\documentclass[aps,prd,10pt,twocolumn,superscriptaddress,nofootinbib,floatfix,
  noshowpacs,preprintnumbers]{revtex4-1}
  \pdfoutput=1
\usepackage{bm}
\usepackage{epsfig}
\usepackage{amsmath}
\usepackage{amssymb}
\usepackage{slashed}
\usepackage{color}
\usepackage[dvipsnames]{xcolor}
\usepackage[colorlinks=true,breaklinks=true]{hyperref}
\hypersetup{allcolors=[rgb]{0.0 0.0 0.6},linkcolor=[rgb]{0.75 0.05 0.05}} 

\renewcommand\({\left(}
\renewcommand\){\right)}
\renewcommand\[{\left[}
\renewcommand\]{\right]}

\newcommand{\exclude}[1]{}

\begin{document}

\preprint{MPP-2018-245}
\preprint{NORDITA-2018-095}

\title{Neutrino mass from bremsstrahlung endpoint in coherent
scattering on nuclei}

\author{Alexander~Millar}
\email[Electronic address: ]{alexander.millar@fysik.su.se}
\affiliation{The Oskar Klein Centre for Cosmoparticle Physics,
Department of Physics,
Stockholm University, AlbaNova, 10691 Stockholm, Sweden}
\affiliation{Nordita, KTH Royal Institute of Technology and
Stockholm
  University, Roslagstullsbacken 23, 10691 Stockholm, Sweden}

\author{Georg~Raffelt}
\email[Electronic address: ]{raffelt@mpp.mpg.de}
\affiliation{Max-Planck-Institut f\"ur Physik
(Werner-Heisenberg-Institut),
  F\"ohringer Ring 6, 80805 M\"unchen, Germany}

\author{Leo~Stodolsky}
\email[Electronic address: ]{les@mpp.mpg.de}
\affiliation{Max-Planck-Institut f\"ur Physik
(Werner-Heisenberg-Institut),
  F\"ohringer Ring 6, 80805 M\"unchen, Germany}

\author{Edoardo~Vitagliano}
\email[Electronic address: ]{edovita@mpp.mpg.de}
\affiliation{Max-Planck-Institut f\"ur Physik
(Werner-Heisenberg-Institut),
  F\"ohringer Ring 6, 80805 M\"unchen, Germany}

\begin{abstract}
  We calculate the coherent bremsstrahlung process $\nu+{\cal N} \to
  {\cal N}+\nu+\gamma$ off a nucleus ${\cal N}$ with the aim of revealing the neutrino
  mass via the photon endpoint spectrum. Unfortunately, the large
  required power of a monochromatic neutrino source and/or large
  detector mass make it difficult to compete with traditional
  electron-spectrum endpoint measurements in nuclear $\beta$ decay.
  Our neutral-current process distinguishes between Dirac and Majorana
 neutrinos, but the change of the photon spectrum is of the order of
  $m_\nu/E_\nu$ and thus very small, despite the final-state neutrino
  coming to rest at the photon endpoint.  So the ``Dirac-Majorana
  confusion theorem'' remains intact even if $E_\nu\gg m_\nu$ applies
  only for the initial state.
\end{abstract}

\maketitle

\section{Introduction}

Two important questions of neutrino physics could be addressed if it
were possible to have an experimental source of ``stopping'' or
``slow'' neutrinos. The first is that of the neutrino mass. If the
neutrino has an energy comparable to its mass, then various phase
space or kinematic factors will vary in a way involving the mass and
so offer a way of determining it.

Secondly, there is the question of whether neutrinos are Majorana or
Dirac. If the neutrino is a self-conjugate particle (Majorana) it has
only an axial current, as opposed to the non-self-conjugate (Dirac)
case where it has both vector and axial currents. According to the
``confusion theorem'' \cite{Kayser:1989iu}, however, both cases lead
to the same experimental results when the neutrino is
relativistic. But with slow or non-relativistic neutrinos, differences
can appear in principle and can offer a way to determine the nature of
the neutrino.

Unfortunately neutrinos are invariably produced with relativistic
energies so that it has been difficult to carry out such studies.  One
line of work, such as the KATRIN experiment~\cite{Wolf:2008hf}, has
been to study $\beta$-decay near the endpoint of the spectrum where
the electron has almost all the energy so that the neutrino is
slow. This approach involves very sensitive and elaborate electron
spectro\-metry. Furthermore, as this is a charged-current process, it
is not sensitive to the Majorana vs\ Dirac question.

A new signature to search for low-mass WIMP dark matter was recently
proposed \cite{Kouvaris:2016afs} in the form of coherent
bremsstrahlung $\chi+{\cal N}\to {\cal N}+\chi+\gamma$.  The photon is emitted by
the very non-relativistic nucleus ${\cal N}$ with a mass taken to be much
larger than that of the WIMP. At the endpoint of the photon spectrum,
all kinetic energy of the WIMP is released as electromagnetic energy
which is relatively easy to detect with good precision. So in the CM
frame, which here is almost the same as the laboratory frame, both the
nucleus and WIMP come fully to rest when the photon energy is maximal.
The crucial point is that the photon spectrum, although at a low
absolute level, continues smoothly all the way to the maximum possible
energy.

Inspired by this idea and by the recent observation of coherent
neutrino scattering \cite{Akimov:2017ade} one may wonder if the same
process, substituting a neutrino~$\nu$ for the WIMP~$\chi$, could be
useful to measure the neutrino mass $m_\nu$ because, near the photon
endpoint spectrum, one produces slow neutrinos.  For a fixed initial
energy $E_\nu$, the photon endpoint energy and the shape of the
endpoint spectrum must depend on~$m_\nu$.   A related idea is to produce the neutrinos themselves through an inelastic process~\cite{Berryman:2018qxn} 

The purpose of this short paper is to calculate the bremsstrahlung
spectrum of this process, which is coherent both for the
neutrino-nucleus and nucleus-photon interaction, with particular
attention to the Dirac vs\ Majorana difference. Moreover, we will
estimate the detection rate for a plausible experimental setup.

\section{Calculation of neutrino induced bremsstrahlung}

\begin{figure}[!t]
\hbox to\columnwidth{\includegraphics[width=4cm]
{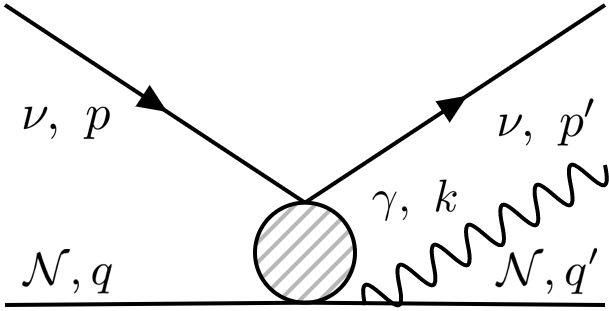}\hfill
\includegraphics[width=4cm]{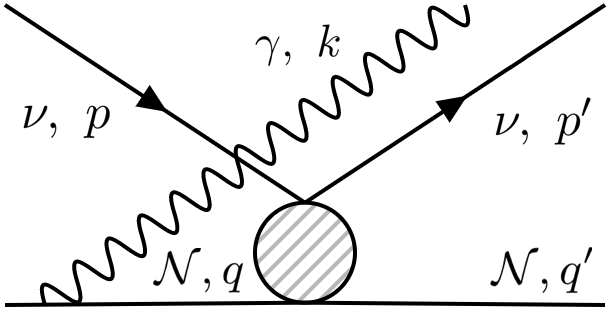}}
\caption{Feynman diagrams for coherent neutrino-nucleus scattering
  with bremsstrahlung. The neutrino is denoted $\nu$ with initial and
  final momenta $p$ and $p'$, the nucleus ${\cal N}$ with $q$ and $q'$, and
  photon $\gamma$ with momentum $k$.}
\label{fig:diagrams}
\end{figure}

We consider a monochromatic neutrino beam scattering off a heavy and
stable nucleus, such as tungsten. The elastic version of this process,
i.e., without photon emission, was calculated long
ago~\cite{Drukier:1983gj}, however due to the relativistic nature of
the outgoing neutrino this process is insensitive to the neutrino
mass. The same cannot be said for the bremsstrahlung process shown in
Fig.~\ref{fig:diagrams} if we consider the regime where almost all of
the available kinetic energy is taken by the photon. We will neglect
the effects of the nucleus being inside an atom because the timescale
of the interaction is much faster than the average timescale for the
electrons to respond~\cite{Kouvaris:2016afs}. 

The bremsstrahlung cross section follows from the Feynman diagrams
shown in Fig.~\ref{fig:diagrams}, where we denote the neutrino initial
and final four momenta $p$ and $p'$, the nucleus four momenta $q$ and
$q'$, and the photon four momentum $k$. The calculation vastly
simplifies in our case where the nuclear mass $M$ is much larger than
the neutrino energy $E_\nu$ or the photon energy $\omega$, allowing us
to ignore recoil effects.  Moreover, in this limit the nucleus
propagator simply contributes $(M\omega)^{-2}$ to the squared matrix
element.  The photon-nucleus vertex contributes a factor
$|\epsilon_\mu \Delta q^\mu|^2$, where $\epsilon$ is the photon
polarization vector and $\Delta q=q-q'$ the four-momentum change of
the charged particle. The photon transversality condition
$\epsilon_\mu k^\mu=0$ implies that we can add or subtract $k$ to
$(q-q')$ so that, after imposing the energy-momentum conservation of the
overall process, $\epsilon_\mu \Delta q^\mu=-\epsilon_\mu \Delta
p^\mu$. In the Coulomb gauge $\epsilon$ has only spatial parts and after
averaging over directions of photon emission and summing over photon
polarisations, the rate is proportional to $|\Delta {\bf
  p}|^2$. Overall we find the bremsstrahlung cross section
\begin{align}
\frac{d^4\sigma_{\rm b}}{d^3{\bf p'}\,d\omega}
={}&\frac{Z^2\alpha}{(2\pi)^3}\frac{1}{12M^4E'_\nu|{\bf p}|}
\frac{|\Delta{\bf p}|^2}{\omega}\,
|{\cal M}_{\rm s}|^2
\nonumber\\[1ex]
&{}\times\delta^{(0)}(E_\nu-E'_\nu-\omega)\,,
\end{align}
where ${\cal M}_{\rm s}$ is the matrix element of $\nu {\cal N}$ scattering
in the large-$M$ limit.

We then integrate over the remaining phase space,
where we use $d^3|{\bf p}'|=
2\pi\, d\cos\theta\,|{\bf p}'|^2 d|{\bf p}'|$. This is used to
integrate over the energy delta function, so a Jacobean
$d|{\bf p}'|=E'_\nu/|{\bf p}'|\,dE'_\nu$ comes in and we find
\begin{equation}\label{eq:softphoton}
\frac{d^2\sigma_{\rm b}}{d\omega\,d\cos\theta}
=\frac{2Z^2\alpha}{3\pi \omega}\,\frac{|\Delta{\bf p}|^2}{M^2}\times
\frac{1}{2\pi}\frac{1}{16M^2}\frac{|{\bf p}'|}{|{\bf p}|}|{\cal M}_{\rm s}|^2\,,
\end{equation}
where $|{\bf p}|=\sqrt{E_\nu^2-m_\nu^2}$ and $|{\bf p}'|=\sqrt{(E_\nu-\omega)^2-m_\nu^2}$.
Moreover, $\theta$ is the
neutrino scattering angle and the second factor is the $\nu {\cal N}$
scattering cross section $d\sigma_{\rm s}/d\cos\theta$, however with
the alteration that $E_\nu'=E_\nu-\omega$ (heavy nucleus
approximation).

In the soft-photon limit ($\omega\ll E_\nu$) this becomes the ordinary
elastic $\nu {\cal N}$ scattering cross section. In this limit we can ignore
${\bf k}$ in the momentum balance so that $|\Delta{\bf p}|=|\Delta{\bf
  q}|$ and therefore $|\Delta {\bf p}|^2/M^2\to |\Delta {\bf v}|^2$,
where $\Delta {\bf v}$ is the velocity change of the nucleus. Our
result then agrees with the usual classical nonrelativistic
soft-photon bremsstrahlung rate, serving as a verification of our
overall factors.

Finally, we need the matrix element $|{\cal M}_{\rm s}|^2$ derived in
Appendix~\ref{appendix}.  Here we assume a Dirac neutrino, discussing
the difference to Majorana in Sec.~\ref{DvsM}. We assume that the
nucleus is scalar, however our results hold for any large nucleus,
such as tungsten, as only the coherently enhanced vector coupling
matters~\cite{Drukier:1983gj}. In addition, we use $M\gg E_\nu\gg
m_\nu$, ignore the recoil of the nucleus, and ignore $m_\nu$ everywhere
except in the final-state neutrino variables. With these
approximations we find
\begin{align}
|{\cal M}_{\rm s}|^2={}&4G_{\rm F}^2M^2
E_\nu\[Z\,(1-4\sin^2\theta_{\rm W})-N\]^2\nonumber \\
&\times\(E_\nu'+|{\bf p}'|\cos\theta\)\,,
\end{align}
where $Z$ is the nuclear charge and $N$ the neutron number.
Henceforth we will neglect
the neutral-current term on protons proportional to
$(1-4\sin^2\theta_{\rm W})=0.075$.
Performing the final phase-space integral to remove the neutrino scattering
angle we find
\begin{align}\label{eq:spectrumNOTendpoint}
\frac{d\sigma_{\rm b}}{d\omega}={}&\frac{G_{\rm F}^2
N^2Z^2\alpha}{6\pi^2M^2\omega}
|{\bf p}'|\[E'_\nu\(|{\bf p}|^2+|{\bf p}'|^2\)-\frac{2|{\bf p}||{\bf p}'|^2}{3}\],
\end{align}
where we may use $|{\bf p}|=E_\nu$ everywhere.

\begin{figure}[!t]  
\centering
\includegraphics[width=8cm,height=4.5cm]{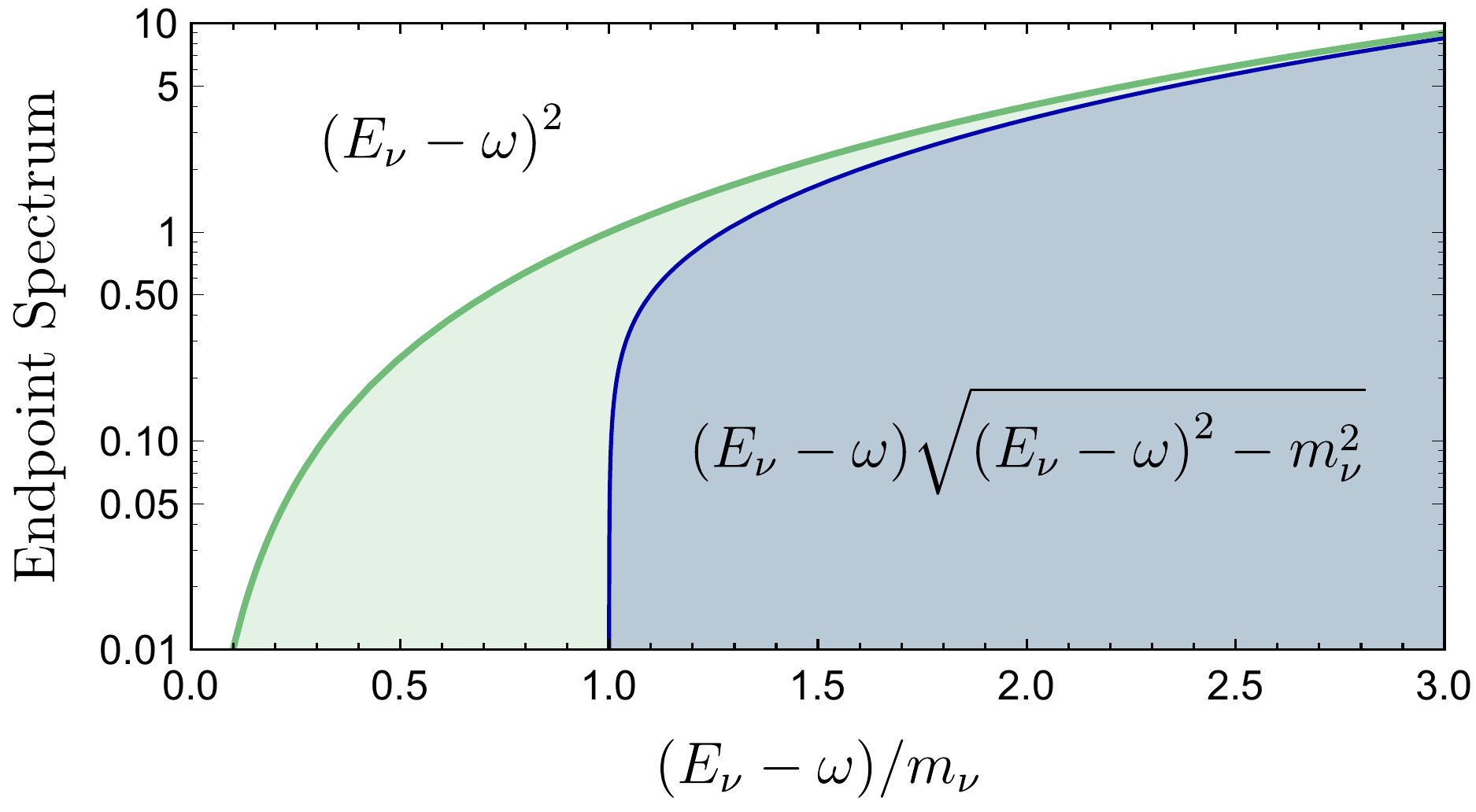}
\caption{Endpoint spectrum of $\nu+{\cal N}\to {\cal N}+\nu+\gamma$ for
  monochromatic incident neutrinos (energy $E_\nu$) with mass $m_\nu$
  (blue) and zero mass (green). There is no difference between
  Majorana and Dirac neutrinos when $E_\nu\gg m_\nu$. $E_\nu-\omega$
  is the energy of the outgoing neutrino.}
\label{fig:endpoint}
\end{figure}

We will be interested in the endpoint of the cross section, i.e.,
when $|{\bf p}'|\ll |{\bf p}|$. In this case we get
\begin{equation}\label{eq:spectrum}
\frac{d\sigma_{\rm b}}{d\omega}=\frac{G_{\rm F}^2
N^2Z^2\alpha}{6\pi^2M^2\omega}E_\nu^2\,E'_\nu|{\bf p}'|\,,
\end{equation}
where
\begin{equation}
E'_\nu|{\bf p}'|=(E_\nu-\omega)\sqrt{(E_\nu-\omega)^2-m_\nu^2}\,.
\end{equation}
This result reveals how one would measure the
neutrino mass. When the outgoing neutrino is relativistic the
cross section scales as ${\mbox (E_\nu-\omega)^2}$, but experiences
a sharp cut-off at $m_\nu$. To see the endpoint explicitly, we plot
both ${\mbox(E_\nu-\omega)^2}$ and ${\mbox
(E_\nu-\omega)\sqrt{(E_\nu-\omega)^2-m_\nu^2}}$ in
Fig.~\ref{fig:endpoint}. As expected, for $(E_\nu-\omega)\lesssim
{\rm few}~m_\nu$ there is a significant difference between the
massive and massless cases.

So far we have considered only a single generation of neutrinos, but
all three mass eigenstates are produced by a neutrino source which, if
produced by charged-current interactions, provides ratios determined
by the mixing angles.  As coherent scattering is via the neutral
current, all three mass states interact equally. Thus one would in
fact see three ``endpoints,'' each causing a dip proportional to the 
production rate.

\section{Dirac vs Majorana neutrinos}
\label{DvsM}

A longstanding question is whether neutrinos are Dirac or Majorana. The difficulty in answering this question is due to the ``Dirac-Majorana confusion theorem".  When the neutrino masses are small compared to the other energy scales in the experiment then the difference between Dirac and Majorana neutrino interactions become suppressed by factors of $m_\nu/E_\nu$~\cite{Kayser:1989iu}. Specifically, the two Majorana helicity states are equivalent to the left handed particle and right handed antiparticle Dirac states when $m_\nu \to 0$. In Ref.~\cite{Kayser:1989iu} the authors noted that the confusion theorem holds even when one of the neutrinos was non-relativistic, however the relative size of the suppression compared to the purely relativistic case was not studied. 
 
At tree level, for experiments studying tritium or other beta decays
\cite{Wolf:2008hf}, there is no difference as the process is via the
charged-current interaction. In this case the Feynman rules and thus
matrix elements are the same~\cite{Gluza:1991wj}.  However, in our
neutral-current process there is a difference between the Dirac and
Majorana squared matrix elements which is found in
Appendix~\ref{appendix} to be
\begin{equation}
\Delta|{\cal M}_{\rm s}|^2\sim 4G_{\rm F}^2M^2N^2m_\nu^2\,.
\end{equation}
For simplicity we have neglected here terms that are higher order in
$m_\nu$ for slow final-state
neutrinos as discussed in Appendix~\ref{appendix}.
The fractional difference is
\begin{equation}
\frac{\Delta|{\cal M}_{\rm s}|^2}{|{\cal M}_{\rm s}|^2}
\sim\frac{m_{\nu}^2}{E_\nu(E'_\nu+|{\bf p}'|\cos\theta)}\,.
\end{equation}
When the final neutrino is non-relativistic, ${\mbox E'_\nu\sim
  m_\nu}$, the fractional difference is ${\cal O}(m_\nu/E_\nu)$. While
for relativistic neutrinos this is ${\cal O}(m_\nu^2/E_\nu^2)$, and in
this sense our slow final-state neutrinos somewhat improve the
situation, the relative difference is still extremely small.   In other words, it is relatively easier to distinguish Dirac and Majorana neutrinos only because the terms which are the same for both Dirac and Majorana neutrinos are suppressed near the kinematic endpoint. 

\begin{figure}[!t]  
\centering
\includegraphics[width=8cm,height=4.5cm]{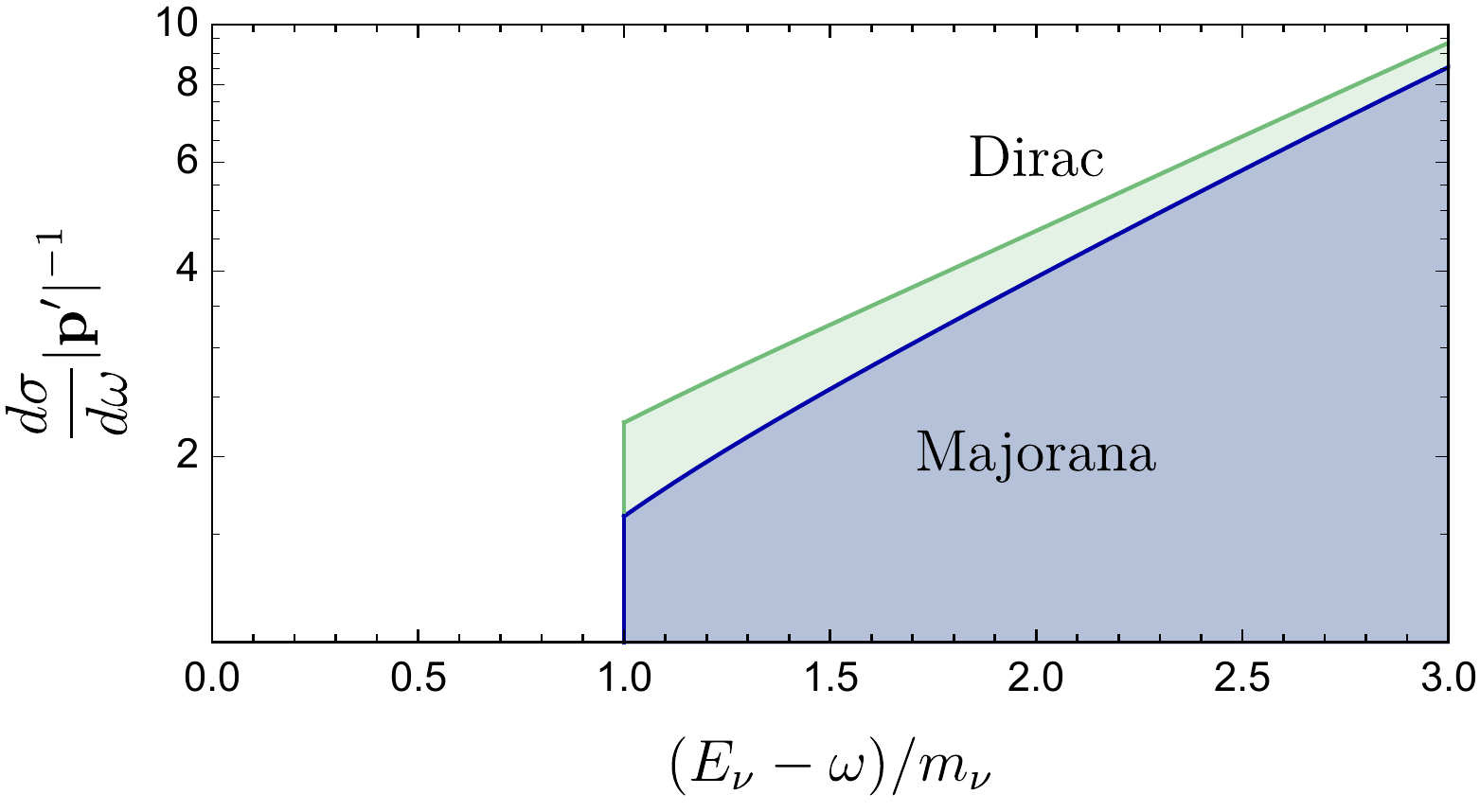}
\caption{Differential cross section $d\sigma/d\omega$ vs the scaled final neutrino energy $(E_\nu-\omega)/m_\nu$ for Dirac (green) and Majorana (blue) neutrinos. We show a moderately relativistic initial neutrino energy \hbox{$E_\nu=3 m_\nu$}. We scale by the neutrino momentum $|{\bf p'}|$ to better show the fractional differences.}
\label{fig:dvsm}
\end{figure}

To illustrate the difference, in Fig.~\ref{fig:dvsm} we plot the differential cross section $d\sigma/d\omega$ for an exaggerated case of $E_\nu=3 m_\nu$. To emphasise the 
behaviour near the endpoint we have divided the predicted spectra by $|{\bf p'}|$, 
the momentum of the outgoing neutrino. We assumed a mixture of neutrino helicities as would be produced from a charged current source, however this is not mandatory.  Although there are quantitative 
effects, it seems that there is little characteristic difference between 
the Majorana and Dirac cases. Such an experiment would require a very strong, low 
energy, source of neutrinos. 
 
\section{Experimental Estimates}

\subsection{Endpoint Cross Section}

Of course, the Dirac vs\ Majorana question only makes sense if the
bremsstrahlung method allows us to find the neutrino mass in the first
place. To this end, the photon spectrum needs to be measured with
sufficient precision in the endpoint region. A minimal requirement is
to measure a significant number of events in a region within a few
$m_\nu$ of the massless endpoint. To arrive at a numerical estimate we
therefore define the endpoint cross section as the integral of the
massless cross section from the endpoint to $2m_\nu$,
\begin{align}\label{eq:crosssection}
\sigma_{\rm end}&{}=\int_{E_\nu-2m_\nu}^{E_\nu}
\frac{d\sigma_{\rm b}}{d\omega}\Big|_{m_\nu=0}d\omega
\nonumber\\
&{}=\frac{4\alpha}{9\pi^2}\(\frac{NZ}{A}\)^2
\frac{G_{\rm F}^2E_\nu m_\nu^3}{m_u^2}
+{\cal O}(m_\nu^4)\,,
\end{align}
where $A$ is the mass number of the nucleus, so $M=A m_u$ with
$m_u=930~{\rm MeV}$ the atomic mass unit.

We see from Eq.~\eqref{eq:crosssection} that by going to heavier
elements we gain by a coherence factor $(NZ/A)^2$. In fact, only the
$\nu {\cal N}$ scattering process provides a significant coherent enhancement
because the bremsstrahlung modification contributes $(Z/A)^2$, where
the factor $Z^2$ comes from the coherent photon interaction with the
nucleus and $1/A^2$ from the decrease of $|\Delta {\bf p}|^2/M^2$ for
a heavier nucleus.

\subsection{Nuclear Recoil Approximation}

Our most significant approximation is the neglect of the kinetic
energy of the nucleus. This enters in deriving
Eq.~\eqref{eq:softphoton} and the above simplifications of the
three-body kinematics.  Negligible nuclear recoil is also required
experimentally as one needs to avoid excessive blurring of the
endpoint spectrum by nucleus recoil. The photon could emerge in any
direction relative to the impinging neutrino, so the nucleus needs to
absorb a momentum between 0 (forward emitted photon) and $|\Delta {\bf
  q}|=2E_\nu$ (backward emission), so the recoil energy $E_{\rm rec}$
is in the range between 0 and $(2 E_\nu)^2/2M=2E_\nu^2/M$. The
condition $E_{\rm rec}<m_\nu$ implies
\begin{equation}\label{eq:3body}
  E_\nu<\sqrt{2Mm_\nu}=590~{\rm
keV}\,\sqrt{\frac{A}{184}\,\frac{m_\nu}{1~{\rm eV}}}\,,
\end{equation}
where $m_\nu$ is the smallest neutrino mass to which the experiment
could be sensitive and the numerical value is for tungsten with an
atomic weight of $A=184$. When this approximation is not valid more detailed calculations are necessary.

\subsection{Estimated Cross Section}

We use tungsten ($Z=74$) as an example as used in the coherent
scattering project ``$\nu$-cleus''~\cite{Strauss:2017cuu}.
Using the largest useful $E_\nu$ of Eq.~(\ref{eq:3body}) as a
benchmark, our endpoint cross section on tungsten is
\begin{equation}\label{eq:supersmall}
\sigma_{\rm end}=2.3\times10^{-68}~{\rm
cm}^2\,\(\frac{m_\nu}{1~{\rm eV}}\)^{3.5}.
\end{equation} 
Evidently a very powerful source of monochromatic neutrinos would be
needed to carry out such a study experimentally.

The most promising source would consist of elements which decay solely
through electron capture. Two possible candidates are $^{68}$Ge and
$^{51}$Cr. Both decay via electron capture, with a lifetime of 270 and
27 days respectively.  Chromium was planned as the source for the SOX
experiment~\cite{Borexino:2013xxa} and was used in the GALLEX
experiment~\cite{Hampel:1997fc}.  To study the Majorana vs Dirac
question the neutrino source would additionally have to be of low
energy so the $m_\nu/E_\nu$ is not entirely negligible. In that case,
Ge with its decay at at 110~keV would be preferable, and would allow
one to probe $m_\nu\sim 0.1$~eV.

Either way, it would be extremely challenging to find a combination of
realistic source power and detector size capable of overcoming the
small cross section of Eq.~(\ref{eq:supersmall}). While experiments
using these types of sources have been performed, the relevant cross
section was an ordinary weak one. In particular for the GALLEX
experiment using a $^{51}$Cr source, the absorption cross section on
gallium was $58\times10^{-46}~{\rm cm}^2$ \cite{Hampel:1997fc}.  In
our case, we need many events in the very narrow energy range near the
endpoint defined by the neutrino search mass, which is what makes
Eq.~(\ref{eq:supersmall}) so small. Moreover, the nature of our
process as a radiative process of the order of $\alpha$ relative to a
weak process and the nonrelativistic nature of the radiation process,
bringing in the factor $|\Delta{\bf p}|^2/M^2\ll 1$, is not completely
overcome by coherence factors.

One possible way to improve the cross section would be to consider an
electron rather than a nucleus as a target to avoid the large $1/M^2$
factor. Unfortunately one loses the coherent enhancement of the cross
section, and must deal with the significantly more complicated
kinematics.  A similar process was considered in \cite{Akhmedov:2018wlf}, though not for the purposes of neutrino mass detection.

\section{Conclusions}

In this paper we have explored the possibility of using coherently
enhanced inelastic neutrino scattering to determine the neutrino
masses. As the photon can carry the entire available kinetic energy of
the system one could measure the masses of the neutrinos by measuring
the endpoint of the photon spectrum.

To this end, we calculated the coherent bremsstrahlung produced by
neutrino scattering for both Dirac and Majorana neutrinos.  We showed
that suppressions to the rate caused by the low velocity of the
nucleus and by moving to the kinematic endpoint strongly reduce the
cross section relative to an ordinary weak one.

Although impractically intense monochromatic neutrino beams would
seem to be needed for an experiment, a number of interesting points
arise from the calculation. Since we go beyond the soft photon
approximation we are able to establish the shape of the photon
spectrum near the endpoint, namely
$E_\nu'|{\bf p}'|=(E_\nu-\omega)\sqrt{(E_\nu-\omega)^2-m_\nu^2}$.

Further, we explored the confusion theorem in this context. While in
principle there is a difference in the rate between Dirac and Majorana
neutrinos, it is suppressed by a factor of $m_\nu/E_\nu$ and so one
would need an exceedingly precise measurement with a low energy source
to detect it.

Finally we would like to note the principle we apply here could be
applied to other reactions, some of which might be more favourable
experimentally. These processes would consist of a quasi-elastic
neutrino scattering, leading to a three-body final state. Taking one
of the particles to its kinematic endpoint can lead to a ``stopping''
final neutrino.

\section*{Acknowledgements}

We acknowledge partial support by the Deutsche Forschungsgemeinschaft
through Grants No.\ EXC 153 (Excellence Cluster “Universe”) and SFB
1258 (Collaborative Research Center ``Neutrinos, Dark Matter,
Messengers'') as well as by the European Union through Grants
No.\ H2020-MSCA-ITN-2015/674896 (Innovative Training Network
``Elusives'') and H2020-MSCA-RISE-2015/690575 (Research and Innovation
Staff Exchange project ``Invisibles Plus''). AM is supported by the
European Research Council under Grant No.\ 742104.

\appendix
\section{Matrix elements}
\label{appendix}

We calculate the matrix element for the scattering process $\nu+{\cal N}\to
{\cal N}+\nu$ of a relativistic Dirac or Majorana neutrino with mass $m_\nu$
and energy $E_\nu\gg m_\nu$ on a heavy nucleus with mass $M\gg E_\nu$.
Using the Feynman rules of Ref.~\cite{Gluza:1991wj} we find
\begin{subequations}\label{eq:matrixelement}
\begin{align}
  {\cal M}_{\rm s}^{\rm D} &=
  \frac{G_{\rm F}}{2\sqrt{2}}\overline\nu_{\rm f}\gamma^\mu(1-\gamma^5)\nu_{\rm
i}J^{\cal N}_\mu\,, &&{\rm Dirac} \\
     {\cal M}_{\rm s}^{\rm M} &=\frac{G_{\rm
F}}{\sqrt{2}}\overline\nu_{\rm f}\gamma^\mu\gamma^5\nu_{\rm
i}J^{{\cal N}}_\mu\,,   &&{\rm Majorana}
\end{align}
\end{subequations}
where we have neglected global phases. Moreover
\begin{equation}
J^{{\cal N}}_\mu=\[(1-4\sin^2\theta_{\rm w})Z-N\](q+q')_\mu
\end{equation}
is the matrix element of the nucleus current
with $N$ the number of neutrons and $Z$ of protons.
We have assumed the nucleus to be scalar
or that it has so many nucleons that we can ignore the axial
interaction relative to the
coherent vector one. 
Note that for the Majorana case we simply needed to replace
$(1-\gamma^5)$ with $2\gamma^5$,
i.e., a Majorana neutrino only couples axially.

 One can actually see from this observation that the confusion theorem must apply even when only one of the 
neutrinos is relativistic~\cite{Kayser:1989iu}. For the 
Dirac case, to get to Eq.~\eqref{eq:matrixelement} we must evaluate the spinor expression $\overline u(p')\gamma_\mu(1-\gamma_5 )u(p)$. For the Majorana case there is only the axial 
current, $2 \overline u(p')(\gamma_\mu\gamma_5 )u(p)$, to evaluate. The two cases will be equivalent only if the vector current term yields the same as the 
axial current term. This will be the case if one of the $u$'s is an 
eigenstate of $\gamma_5$, that is, a helicity state. In particular, only one of the $u$'s needs to be a helicity state since the $\gamma_5$ can be passed 
from one side to the other with merely a sign change. Thus in the relativistic limit the difference between Dirac and Majorana will always be suppressed. To see the form more explicitly, however, we must preform the full calculation.

We consider the nucleus to be highly non-relativistic and recoil
effects can be neglected, implying that we can use $J^{\cal 
N}=-N\,(2M,{\bf 0})$,
where we neglect the contribution from protons as explained in the main text. Thus we only
need the 0 component of the neutrino current. Since we will
eventually consider the case $E_\nu\gg m_\nu$  the initial neutrino
source will have an almost definite helicity. In the squared
matrix element we find, assuming an initial helicity $h$ and 
summing over the final one,
\begin{subequations}
\begin{align}
\left|\overline\nu_{\rm f}\gamma^0(1-\gamma^5)\nu_{\rm
i}\right|^2={}&8\left(E_\nu -h|{\bf p}|\right) \left(E_\nu '-h|{\bf
p'}| \cos\theta \right),\\
\left|2\,\overline\nu_{\rm f}\gamma^0\gamma^5\nu_{\rm
i}\right|^2={}&
16\left(E_\nu E_\nu '+|{\bf p}||{\bf p'}|\cos\theta-m_\nu^2\right),
\label{eq:Majorana-squared}
\end{align}
\end{subequations}
where $\theta$ is the neutrino scattering angle, $E_\nu =\sqrt{{\bf
p}^2+m_\nu^2}$
and analogous for $E_\nu '$.  To write the above we
used~\cite{Long:2014zva}
\begin{equation}
\nu \overline \nu=\frac 1 2 (\slashed
p+m_\nu)\left(1+h\gamma^5\slashed S \right)
\end{equation}
with the spin vector
\begin{equation}
S=\left(\frac{|{\bf p}|}{m_\nu}\, ,\, \frac{E_\nu}{m_\nu}\hat{\bf
p}\right)\,.
\end{equation} 
Note that in the Majorana case the result of
Eq.~(\ref{eq:Majorana-squared}) does not depend on the initial
helicity $h$. This is because in the relativistic limit the two
helicities play the role of neutrino and antineutrino and the vector
interaction rate of the neutral current is the same for $\nu$ and
$\bar\nu$. Including axial-current interactions with nuclei, the $\nu$
and $\bar\nu$ scattering rates would be different due to weak
magnetism \cite{Horowitz:2001xf}.  In the Dirac case, for $h=-1$, the
result is the same as for Majorana neutrinos up to $m_\nu^2$
corrections, whereas for $h=+1$ it is much smaller and vanishes for
$m_\nu=0$ because this ``wrong-helicity'' case represents the sterile
right-handed Dirac component. 
%

For non-vanishing $m_\nu$, neutrinos from the source have a
``wrong-helicity'' component with a probability
$(m_\nu/2E_\nu)^2$, which should be included in the rate. For a
nice demonstration of this point, see Appendix~A of
Ref.~\cite{Langacker:1998pv}. Thus there are in principle three
effects that can distinguish between Dirac and Majorana neutrinos.
(i)~The higher probability of $h=+1$ Majorana neutrinos to
interact. (ii)~The difference between $E_\nu$ and $|{\bf p}|$,
which enters via different $S$ dependence of Dirac and Majorana
neutrinos. (iii)~The extra term $m_\nu^2$ in the Majorana case
which tends to cancel the energy terms for small energies. In the elastic case, this term decouples the neutrino from the neutral current at low energies.

Putting this all together with $1-(m_\nu/2E_\nu)^2$ as the
probability
for the $h=-1$ initial state and $(m_\nu/2E_\nu)^2$ for the
$h=+1$ one, we find for the summed squared matrix elements and the
modified flux factors
\begin{subequations}
\begin{align}
\left|{\cal M}_{\rm s}^{\rm D}\right|^2={}&2G_{\rm F}^2{
N}^2M^2\left(1-\frac{m_\nu^2}{4E_\nu^2}\right)
\nonumber\\
&{}\times
(E_\nu +|{\bf p}|)\left (E_\nu '+|{\bf p}'|\cos\theta\right),\\[1ex]
\left|{\cal M}_{\rm s}^{\rm M}\right|^2={}&4G_{\rm F}^2{
N}^2M^2
(E_\nu E_\nu'+|{\bf p}||{\bf p}'|\cos\theta -m_\nu^2)\, .
\end{align}
\end{subequations}
In the Dirac case the correction due to the $h=+1$ component is of
order $(m_\nu/2E_\nu)^4$
through the flux factor $(m_\nu/2E_\nu)^2$ and the cancellation in
$(E-|{\bf p}|)$, and has been left out. Thus these relations are
exact up to ${\cal O}(m_\nu^2)$.
We will consider the case where $E_\nu'\ll E_\nu $ so we can simplify these expressions
by neglecting the wrong-helicity neutrinos and setting $E_\nu
=|{\bf p}|$ and find
\begin{subequations}
\begin{align}
\kern-0.5em\left|{\cal M}_{\rm s}^{\rm D}\right|^2={}&4G_{\rm F}^2{ N}^2M^2
 (E_\nu E_\nu'+E_\nu|{\bf p}'|\cos\theta)\,,\\ 
\kern-0.5em\left|{\cal M}_{\rm s}^{\rm M}\right|^2={}&4G_{\rm F}^2{ N}^2M^2
(E_\nu E_\nu '+E_\nu |{\bf p}'|\cos\theta -m_\nu^2)\,.
\end{align}
\end{subequations}
These are the expressions that are relevant for the endpoint
spectrum discussed in the main text.


\end{document}